# VERSATILE DATA ACQUISITION AND CONTROLS FOR EPICS USING VME-BASED FPGAS[*]

T. Allison, R. Flood, TJNAF, Newport News, VA 23606, USA

## Abstract

Field-Programmable Gate Arrays (FPGAs) have provided Thomas Jefferson National Accelerator Facility (Jefferson Lab) with versatile VME-based data acquisition and control interfaces with minimal development times. FPGAs have been used to interface with VME controllers using standard A16 and A24 address modes. VME vectored-interrupt capability has also been implemented in some applications. FPGA designs have additionally been used to provide control logic for numerous systems by interfacing with Analog to Digital Converters (ADC), Digital to Analog Converters (DAC), various interlocks, and other drive signals. The building blocks of these logic designs can be tailored to the individual needs of each system and provide system operators with read-backs and controls via a VME interface to an EPICS based computer. This versatility allows the system developer to choose components and define operating parameters and options that are not readily available commercially. Jefferson Lab has begun developing standard FPGA libraries that result in quick turn around times and inexpensive designs. There have been approximately twelve VME-based FPGA designs implemented in the RF and Electronic Support (RFES) Group at Jefferson Lab. FPGA logic density and device speed continue to increase which enables many system designs to be incorporated onto one FPGA. FPGA designs can manipulate data quickly due to the small processing overhead associated with a custom design. This coupled with physical performance advances and optimized logic from compiler tools makes FPGAs solutions faster than many microprocessors. The ability to modify and simulate this firmware enables a designer to easily add new enhancements to a system or modify existing parameters, permitting the design to be both flexible and expandable for future applications.

## 1 INTRODUCTION

In the early years of VME design, interfaces typically required a substantial number of discrete TTL devices and Printed Circuit Board (PCB) space to decode and generate the necessary timing signals for proper VME communication. The interface was hard-wired leaving little ability to change or add features to the module after committing the design to fabrication. Commercial modules were typically expensive and often did not provide the specific features required by unique laboratory applications.

With the introduction of fusible-link logic and eventually, ultraviolet erasable and electrically erasable Programmable Logic Devices (PLDs), VME interface designs began to take on a more generic appearance. PLDs contained the application-specific logic, such as control logic, address decoding, and register decoding which defined the board's function. The ability to reprogram the firmware allowed for logic corrections to be made to the design even after board fabrication. Increased use of Computer Aided Design (CAD) tools permitted generic designs to easily be re-used or enhanced for new applications.

The Altera 7000 series FPGA devices were among the first FPGAs to be incorporated in VME designs at Jefferson Lab. These devices provided much greater logic densities than first-generation Electrically Programmable Logic Devices (EPLDs), often at lower cost and in smaller packages. In addition to incorporating a simple VME slave interface, several registers and complex state-machines could now be implemented directly in the device. Component size was also changing with smaller Plastic Leadless Chip Carrier (PLCC) and Quad Flat Pack (QFP) packaging becoming more commonplace. Most members of this logic family still required removal from the circuit board for re-programming.

Recent VME and stand-alone designs at Jefferson Lab have used Altera FLEX10K or ACEX1K devices. These are SRAM-based components requiring a companion Electronically Erasable Programmable Read Only Memory (EEPROM) configuration device. All of theses components are available in a variety of packages, including PLCC, QFP, and Ball Grid Array (BGA). Device voltages of 5V, 3.3V, and 2.5V have been utilized, depending on the logic family chosen and overall project design. All Altera SRAM-based and EEPROM devices include a Joint Test Action Group (JTAG) interface. JTAG is a boundary scan technology used for board level interconnection integrity testing but also provides In System Programming (ISP) capability. This allowed for easy reprogramming of the FPGA without having to remove the component or use socket devices.

---

[*]This work was supported by the U.S. DOE Contract No DE-AC05-84-ER40150

Many applications have been embedded into FPGAs since their introduction at Jefferson Lab. Data acquisition and system controls have been predominating. More recent applications have been performing mathematical operation and will expand to DSP algorithms and embedded processors.

## 2 TOOLS

The logic for the FPGAs is mainly developed using Altera Hardware Description Language (AHDL) with Altera's Maxplus II software. This is a specialized hardware description language that is optimized for Altera devices. It is largely based on Very High Speed Integrated Circuit (VHSIC) Hardware Description Language (VHDL) but targets the timing characteristics and physical structure of Altera devices to provide accurate timing simulations and help the designer to determine the correct Altera FPGA for an application.

Once the generic VME interface was constructed in AHDL, the code was reused in multiple designs with slight modifications for each application. This is also the case when communicating to DACs, ADCs, SDRAM, FLASH memory, and any other device. This reuse allows for short development times and can be molded to the needs of each system.

VHDL has also been used in some designs with the ALDEC Active-HDL software. The ALDEC compiler does not target a device like the Altera compiler so the simulations do not incorporate the timing characteristics of the FPGA. This compiler does allow the designer to fit a design to almost any FPGA from any manufacturer however only Altera FPGAs have been implemented in the RFES group at Jefferson Lab.

P-CAD is the PCB design tool used to develop boards at Jefferson Lab. Once the VME interface is defined in a schematic and on a PCB layout then it can also be reused to cut down on development time. The most popular FPGAs used at Jefferson Lab are from the Altera FLEX10K series. This family was chosen because they are available in a 240-pin QFP that does not take up a large amount of board space but allows for many input and output pins. The devices range for 20,000 gates to 100,000 gates and come in 5 volt, 3.3 volt, and 2.5 volt packages. This allows for the same PCB footprint to be used as well and the same logic code. Other devices have also been used and have similar characteristics.?

## 3 APPLICATIONS

### 3.1 System Catch All Module (SCAM)

The SCAM was developed as a replacement for an existing hard-wired VME module that had been in service for a number of years. The SCAM module generates timing pulses used to modulate the three-beam laser used in the polarized source of the Continuous Electron Beam Accelerator Facility (CEBAF) injector at Jefferson Lab. Thus supporting independent beam delivery to each of the three experimental halls.

The PCB is a 4-layer 6U module with an 8-bit VME slave interface. The front panel supports 16 optical input/outputs as well as 18 Lemo connectors. A GAL20V8 provides VME A16 address-decoding for seven 8-bit registers contained in an 84-pin Altera EPM7160 EPLD operating at 16Mhz. Numerous state-machines and counters were implemented in the device.

### 3.2 Injector High Voltage Controller

The Injector High Voltage Controller provides control of four high-voltage relays and a 100kV power supply used to power the electron guns in the CEBAF injector. Two voltage-to-frequency converters and two 16-bit DACs are used to set and read back the power supply voltage. Many interlocks that shut off the power supply are incorporated to ensure that personnel are not exposed to high voltage. The module also slowly ramps the voltage to the set point and provides an over-current limit to prevent damaging the electron guns. When the power supply is turned off, a timer and a voltage read-back is used to ensure the high voltage bleeds off before it can be turned back on. This helps avoid arcing.

The PCB is a 4-layer 6U board with an 8-bit VME slave interface. The front panel supports 2 optical outputs as well as a JTAG connector. A GAL20V8 provides VME A16 address-decoding for nine 8-bit registers contained in a 240-pin Altera FLEX10K50 FPGA operating at 10Mhz. The device is re-programmable via the front-panel JTAG interface.

### 3.3 30Hz Board

The 30Hz Board provides synchronous timing signals to various Input/Output Computers in VME crates located throughout the CEBAF accelerator [1]. Communications are established via fiber optic cable in a token ring arrangement. Each board is assigned a unique serial address via on-board jumpers. Address 0 is reserved as a broadcast address in which all boards decode the serial message. Upon receipt of such a message, each board decodes the data and generates a VME interrupt if enabled.

The PCB is a 4-layer 3U board with a 16-bit VME slave interface, which includes D08 vectored-interrupt capability. The A24 address decoding, the VME interface, twelve 8-bit registers, and the serial message encoding/decoding are contained in a 2.5 volt 144-pin Altera ACEX EP1K50 FPGA operating at 20 Mhz.

The FPGA is re-programmable via a front-panel JTAG interface.

### 3.4 Dual DSP Board

The Dual DSP Board is a versatile general-purpose digital signal processing board utilizing two Texas Instrument TMS320C6711 floating-point signal processors [2]. An FGPA is used to provide arbitration between the two DSPs and 128 kilobytes of dual port memory. The FPGA also maps the dual port memory to VME address space and provides communication across a custom P2 back plane.

The module is a 10-layer 6U board and uses an Altera FLEX10K50 FPGA operating at 25 Mhz that supports VME A24 address decoding, a 16-bit VME slave interface, block-transfer cycles, read-modify write cycles, address pipelining, and D08 and D16 vectored-interrupts. Programming of the DSPs and the Altera FPGA is accomplished via two separate front-panel JTAG interfaces.

### 3.5 Machine Protection System (MPS) Comparator

The BCM Comparator module is used to receive and manipulate data from eight of the Dual DSP Boards over a custom P2 back plane. In this application the Dual DSP Boards are configured to measure the beam current in the CEBAF injector and at the multiple end stations. The MPS Comparator module computes beam loss by tallying the end-station currents and comparing the sum to the measured injector current. This difference is the instantaneous loss. The module also compares the current from each location to maximum limits set by the operators. If these limits are exceeded then CEBAF is shutdown. The instantaneous loss is integrated in an adaptive algorithm to produce the integrated loss. If the integrated loss becomes large enough then CEBAF is shutdown. Information about the beam loss is stored in an 8 megabyte SDRAM circular buffer. The SDRAM is mapped to VME address space with a pointer to the start of the buffer. If CEBAF is shutdown then the buffer can be read and the beam loss history reconstructed. A DAC is also used to represent the instantaneous loss and loss data is sent directly to the Machine Control Center via a fiber optic interface.

The module is a 4-layer 6U board and uses a 3.3 volt Altera FLEX10K100 FPGA operating at 40Mhz that supports VME A24 address decoding and a 16-bit VME slave interface. Programming of the Altera FPGA is accomplished via a front-panel JTAG interface.

### 3.6 Phase Lock Loop (PLL) Module

The PLL Module uses a voltage-controlled crystal oscillator to phase lock to a high-Q super-conducting RF cavity, which is then used for testing and commissioning. The operators provide amplitude and phase information to the FPGA through either a remote VME interface or local optical encoders. The FPGA uses the phase and amplitude information to derive in-phase (I) and quadrature (Q) values. I and Q are calculated by multiplying the amplitude times the sine (or cosine respectively) of the phase. According to the input values, the FPGA looks up the corresponding sine, cosine, and amplitude values in FLASH memory and then performs the appropriate multiplications. The resulting I and Q values are then presented to the PLL circuit through two 14-bit DACs. An eight channel ADC is also connected to the FPGA to provide read-back voltages throughout the PLL circuit.

The module is a 4-layer 6U board and uses an Altera FLEX10K50 FPGA operating at 10Mhz that supports VME A24 address decoding and a 16-bit VME slave interface. Programming of the Altera FPGA is accomplished via a front-panel JTAG interface.

## 2 CONCLUSIONS

FPGAs have become an indispensable component in VME board design at Jefferson Lab. Their small size and high logic density make them ideal for VME based data acquisition and controls. The ability to re-configure FPGAs through a JTAG interface provides the designer the ability to incorporate new features into existing design after fabrication and installation. The use of FPGAs and associated embedded logic will continue to be integrated into future designs as they develop. FPGAs will play a major role in the development of the new Low Level RF systems at Jefferson Lab. Digital down-conversion techniques and DSP filtering algorithms are ideal for embedding in devices such as FPGAs. Newer FPGA devices are becoming available with embedded processors surrounded with dense high-speed logic. These advances in technology will make FPGA applications almost limitless.